\documentclass[aps,twocolumn]{revtex4}
\usepackage{amsmath}
\usepackage{amssymb}
\usepackage{graphicx}
\def\bra{\langle}
\def\ket{\rangle}

\begin{document}

\title{The minimal time of dynamic evolution to an arbitrary state}

\thanks{Supported by National Natural Science Foundation of China(Grant No.
10805030) and Major State Basic Research Developing Program 2007CB815005}

\author{Di Lv$^1$, Yan-Song Li$^1$ and Gui-Lu Long$^{1,2}$}
\address{$^{1}$ Key Laboratory for Atomic and Molecular Nanosciences and Department of physics, Tsinghua university, Beijing, 100084\\
$^2$Tsinghua National Laboratory for Information Science and
Technology, Beijing 100084, P R China}

\begin{abstract}
Two bounds on the minimal time of dynamic rotating an initial state
by arbitrary angles have been obtained.  These bounds have been
applied to study the evolutions in the Hadamard-Walsch gate, the
Control-NOT quantum gate,  and the Grover algorithm.
\end{abstract}
\maketitle
%\section{Introduction}

{\bf Introduction.} In the realm of quantum evolution, an important
question is to know the time that an operation needs, i.e., how fast
is the operation. The similar question in classical physics has been
almost solved, but in quantum physics it remains a puzzling problem.
In 1998, Margolus and Levitin \cite{Margolus} have proved that the
shortest time $\tau$ a quantum state takes to its  orthogonal state
is bounded by the inequality (the MV bound hereafter)
\begin{equation}  \label{eq2}
\tau \bar{E}\ge \frac{h}{4},
\end{equation}
where $\bar{E}$ denotes the arithmetic average energy $\bar
E=\langle \psi\vert H\vert \psi\rangle$ of an arbitrary quantum
state $\vert\psi\rangle$ in a given system with  Hamiltonian $H$. It
should be noticed that
 the minimal energy of this  quantum system is set to 0.
 On the other hand, Fleming \cite{Fleming}, Anandan and Aharanov \cite{Anandan}
 and Vaidman \cite{Vaidman} have shown separately that the shortest time needed to
 orthogonalize a quantum state is bounded by another inequality
 (referred to as FAAV bound hereafter),
\begin{equation}  \label{eq1}
\tau\Delta E\ge \frac{h}{4},
\end{equation}
where $\Delta E$ denotes the standard error of energy $\Delta
E=\sqrt{\langle \psi |(H^{2}-{\bar E}^{2})|\psi \rangle}$.

Meanwhile, the shortest time of  applying an outer operation on a
system has been discussed in  \cite{bi11, bi12}. The MV bound and
FAAV bound have been derived in some entangled and non-entangled
systems \cite{bi8,bi9, bi10}. In addition,
  a kind of equations
 that is equivalent to the MV bound in Eq. (\ref{eq1}) and the FAAV bound in Eq. ({\ref{eq2})  have been established
  \cite{bi13,bi14,bi15}. Zielinski and Zych have made a generalization
 from
 which one could derive more details about $\tau$, using energy moments \cite{bi16}. The problem has attracted much attention recently and
 many related problems
 have been studied \cite{bi17,bi18,bi19,bi20,bi21,bi22,bi23,bi24,bi25,bi26,bi27}. Levitin and Toffoli \cite{Levitin} recently  found
 that the MV bound and FAAV bound are tight, and both are satisfied by a quantum state
 that has the property $\bar{E}=\Delta E$. They also discussed what would happen when $\bar{E}\neq \Delta E$.

However, the shortest time that  a state takes to evolve to an
arbitrary state is more general and frequently met in practice. Many
questions remain to be answered.
 For example, does it satisfy the same inequality? If not, is there a new constraint? It is essential to answer these
 questions and to develop results for this more general case and still retaining
 the results in (\ref{eq2}) and (\ref{eq1}) when for evolution to an orthogonal state.
 Estimating the evolution time to an arbitrary state is
 important to study how fast a quantum computer can run, because a quantum computer has a great deal of operations
 that
 changes
  a state to various states. Giovannetti, Lloyd, and Maccone studied this problem and gave the following bound (the GLM-$\alpha$ and GLM-$\beta$
  bounds
  hereafter) \cite{bi8},
\begin{eqnarray}
{\rm max}\left(\alpha(\epsilon){\pi\hbar \over
2E},\beta(\epsilon){\pi\hbar \over 2\Delta   E}\right),
\end{eqnarray}
where $\epsilon=F(\rho,\rho(t))$ is the fidelity between the initial
and final state and
\begin{eqnarray}
\beta(\epsilon)={2\over \pi}\arccos(\sqrt{\epsilon}),
\end{eqnarray}
and $\alpha(\epsilon)$ is determined by a sets of equations.

In this paper, we  derive three bounds for the problem with a
different approach. The results are simple and intuitive.
  Then we apply the bounds to study the evolutions of two quantum gates and the Grover's quantum searching algorithm \cite{Grover}.

%\section{Theory}
{\bf The GLM-$\beta$ bound.} We assume that a given quantum system
evolves from an initial state $\vert\psi(0)\rangle$ into
$\vert\psi(\tau)\rangle$ after time $\tau$ governed by the
Schr\"{o}dinger equation
\begin{equation}\label{schr}
i\hbar\frac{\partial }{\partial t}\vert\psi(t)\rangle =
H\vert\psi(t)\rangle,
\end{equation}
where $H$ is the Hamiltonian of the system. Denote
$\theta=\theta(0\to\tau)\in[0,\;\pi/2]$, which satisfies
\begin{equation}\label{eqtheta}
\cos\theta(0\to\tau)=\vert\langle\psi(0)\vert\psi(\tau)\rangle\vert,
\end{equation}
then we have the first bound
\begin{equation}\label{neqa}
\Delta E \cdot \tau \ge \theta \hbar.
\end{equation}

{\bf The mean-energy bounds.}  Denoting $\bar E$ the average energy
of the system having a time-independent Hamiltonian $H$ and a
minimal energy 0, then one has the {\bf mean-energy bound} (Mean-E
bound) as follows
\begin{equation}\label{neqb}
\bar E \cdot \tau \ge \frac h4\left[1-\sqrt{1+\dfrac4{\pi^2}}\cos\theta\right].
\end{equation}
Now we derive the Mean-E bound. The proof is similar to the one by
Margolus  and Levitin \cite{Margolus}. Expanding the state
$\vert\psi(t)\rangle$ in terms of the energy eigenstates
$\vert\psi_n\rangle$ with nonnegative eigenvalues $E_n$, we obtain
\begin{equation}
\vert\psi(t)\rangle=\sum_{n=1}^\infty c_n \exp\left(-i\frac{E_n}\hbar t\right)\vert\psi_n\rangle,
\end{equation}
where the coefficients $\{c_n\}$ are complex constants satisfying $\sum\limits_{n=1}^\infty \vert c_n\vert^2=1$.
Let
\begin{equation}\label{eqS}
S(\tau)=\langle\psi(0)\vert\psi(\tau)\rangle=\sum_{n=1}^\infty |c_{n}|^{2}\exp\left(-i\frac{E_n}\hbar \tau\right),
\end{equation}
and use the inequality
\[
\cos x \ge 1 - \dfrac{2}{\pi}(x+\sin x),
\]
one can derive
\begin{eqnarray}
\mathrm{Re} S(\tau)&\ge & \sum_{n=1}^\infty \left(1-\frac{2 E_n
\tau}{\pi\hbar}-\frac2\pi\sin\frac{E_n \tau}{\hbar}\right) \vert
c_n\vert^2 \nonumber\\
&=& 1-\frac{4\bar E\tau}{h} + \frac2\pi\mathrm{Im} S(\tau),
\end{eqnarray}
i.e.
\begin{equation}\label{eqRIS}
\bar E\cdot \tau \ge \frac h4 \left[1-\left(\mathrm{Re} S(\tau)-\frac2\pi\mathrm{Im} S(\tau)\right)\right].
\end{equation}
In general, $S(\tau)$ has the form $S(\tau)=\cos\theta\exp(i\varphi)$ and $\varphi$ is a complex phase depending on the evolution time $\tau$.
Thus,
\begin{eqnarray}
\mathrm{Re} S(\tau)-\frac2\pi\mathrm{Im} S(\tau) &=&
\cos\theta\left(\cos\varphi-\frac2\pi\sin\varphi\right)\nonumber\\
& \le &\sqrt{1+\frac4{\pi^2}}\cos\theta,
\end{eqnarray}
from which and inequality (\ref{eqRIS}) one can obtain bound (\ref{neqb}).

If the system has a minimal energy $E_{\rm min} \ne 0$, the bound
will be replaced with the Mean-Min-E bound
\begin{equation}\label{neqb1}
(\bar E-E_{\rm min}) \cdot \tau \ge \frac h4\left[1-\sqrt{1+\dfrac4{\pi^2}}\cos\theta\right].
\end{equation}
If the system has a maximum energy $E_{\rm max}$, it is easy to
prove a similar bound, the Max-Mean-E bound,
\begin{equation}\label{neqb2}
(E_{\rm max}-\bar E) \cdot \tau \ge \frac h4\left[1-\sqrt{1+\dfrac4{\pi^2}}\cos\theta\right].
\end{equation}
Furthermore, if the system has both maximum and minimal energies, it
is obvious
\begin{equation}
\min \{E_{\rm max}-\bar E, \bar E - E_{\rm min}\} \le \delta E\equiv \frac{E_{\rm max}-E_{\rm min}}{2},
\end{equation}
where $\delta E$ is the half-width of energy.  One obtain  a new
bound, the Max-Min bound
\begin{equation}\label{eqdel}
\delta E\cdot \tau \ge \frac h4\left[1-\sqrt{1+\dfrac4{\pi^2}}\cos\theta\right].
\end{equation}

{\bf A tighter bound.} If we use another inequality
\begin{equation}\label{neqabc}
\cos x \ge c-b x+a\sin x
\end{equation}
it can be derived that
\begin{eqnarray}
\mathrm{Re} S(\tau)&\ge & \sum_{n=1}^\infty \left(c-\frac{b E_n
\tau}{\hbar}-a\sin\frac{E_n \tau}{\hbar}\right) \vert
c_n\vert^2 \nonumber\\
&=& c-\frac{b\bar E\tau}{\hbar} + a\mathrm{Im} S(\tau),
\end{eqnarray}

If we choose $a$, $b$, and $c$ properly, when $x>0$, there would be
only one $x$ satisfies the ``='' in (\ref{neqabc}) and others
satisfy ``$>$'', and this process will make the bound tight. Now
assuming $a$, $b$ and $c$ are chosen as above, there should be a
relation $a=a(b,c)$, so it can be derived
\begin{equation}
\bar{E}\tau\ge \frac{\hbar}{b}(c-\sqrt{1+a(b,c)^2}\cos{\theta}).
\end{equation}
Thus we have a tighter bound by maximizing over two parameters, the
BC bound,
\begin{equation}\label{neqd}
\bar{E}\tau\ge
\max_{b,c}{(\frac{\hbar}{b}(c-\sqrt{1+a(b,c)^2}\cos{\theta}))}.
\end{equation}
Usually one has to compute the BC bound numerically. However, to a
good extent, the above bound can be approximated as
\begin{equation}
\frac{\bar{E}\tau}{\hbar}\ge 1.57 - 1.847 \cos{\theta} +0.372
\cos^2{\theta}-0.0958 \cos^3{\theta}.\label{betterapprox}
\end{equation}
It is a good estimation, and has only maximally
$0.0005\frac{\bar{E}\tau}{\hbar}$ difference from the reality.

Compared to the GLM-$\beta$ bound in Ref.\cite{bi8}, when
$\theta=\frac{\pi}{3}$, the BC bound is about $4.5\%$ tighter; and
about $8\%$ tighter when $\theta=\frac{5\pi}{12}$. Fig. \ref{f1}
shows the comparisons of the three bounds: the GLM-$\beta$ bound,
the Mean-E bound and the BC bound. It is shown that the BC bound is
always the tightest. When  $\sin{\theta}$ is small
($\cos{\theta}<0.395$, approximately), the Mean-E bound is tighter
than the GM-$\beta$ bound. However for small evolution, the Mean-E
bound is no longer valid.

\begin{figure}[h]
\begin{center}
\includegraphics{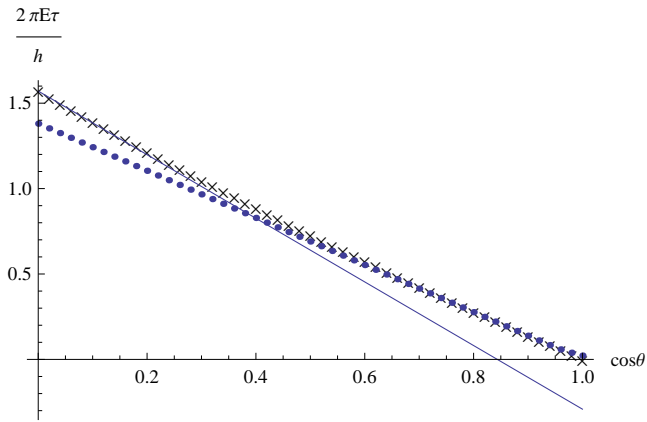}
\caption{Comparisons among the various bounds: ``$\times$" denotes
the BC bound in Eq. (\ref{neqd}), the line denotes the Mean-Min-E
bound in (\ref{neqb1}) and the dots denote the GLM-$\alpha$ bound in
Ref. \cite{bi8}. The BC bound in (\ref{neqd}) is always better than
the GLM-$\alpha$ bound. Meanwhile, the Mean-Min-E bound in
(\ref{neqb1}) is a good estimation when $\theta$ is near
$\frac{\pi}{2}$, but not good around 0. The Mean-Min-E bound in
(\ref{neqb1}) is better than GLM-$\alpha$ bound when $\cos{\theta}$
is small ($\cos{\theta}<0.395$, approximately).}\label{f1}
\end{center}
\end{figure}

%\section{Discussion}
{\bf Discussions.}  One can find that bound (\ref{neqa}) is similar
to bound. When $\theta\to\pi/2$, the GLM-$\beta$
 bound in (\ref{neqa}) and the Mean-E bound in (\ref{neqb}) approach to the MV bound in (\ref{eq1})
 and FAAV bound in (\ref{eq2}) respectively.
 However the Mean-E bound(\ref{neqb}) is not a good
  bound when  $\theta$ is small,  it even has a minus lower limit and the bound becomes
  trivial. However the BC bound is a tighter bound.

It is easy to show that bound (\ref{neqa}) is the tightest when
$\tau \to \mathrm d t$ and $\theta\to\theta(0\to \mathrm d t)$ for
an infinitesimal evolution, because the bound inequality tends to
equality (\ref{eqader}). So there exists no bound such as $\bar
 E\cdot \tau \ge \theta \hbar$ unless $\bar E \ge \Delta E$. In fact, because $1-\sqrt{1+\frac4{\pi^2}}\cos\theta\le1
 -\cos\theta\le2\theta/\pi$ for any $\theta\in[0, \pi/2]$, the Mean-E bound (\ref{neqb}) is weaker than the GLM-$\beta$
 bound in
 (\ref{neqa}) when $\theta\ne \pi/2$
 unless $\bar E< \Delta E$.

The Max-Min bound in (\ref{eqdel}) looks useful when one does not
know details about a system except the width  of energy spectrum.
Indeed the GLM-$\beta$ bound
 (\ref{neqa}) with $\Delta E$ substituted by $\delta E$, i.e.
\begin{equation}\label{neqc}
\delta E\cdot\tau\ge\theta\hbar,
\end{equation}
is the better choice, because one can simply prove $\delta E \ge
\Delta E$.

The GLM-$\beta$ bound in (\ref{neqa}) and the Mean-E bound in
(\ref{neqb}) work independently on the same quantum system, and each
of them confines $\tau$ to
 some extent. But in general the lower limit can not be reached at the same time. An important question arises: Can these bounds
 be attained so that there are not any stronger
inequalities? Levitin and Toffoli  \cite{Levitin} have proved that
there exist states by which bound inequality (\ref{eq1}) or
(\ref{eq2}) (the tighter one) can be asymptotically approached
arbitrarily. It is obvious that the lower limit of (\ref{neqb})
cannot be reached when $\theta$ is small. On the other hand, the
two-level state $(\vert0\rangle+\vert E\rangle)/\sqrt2$ with the
superposition of the ground state $\vert0\rangle$ and the eigenstate
$\vert E\rangle$ of energy $E$ enables $\tau=\theta\hbar/\Delta E$
at any time, where $\Delta E=\delta E=\bar E=E/2$. In general, it
cannot be ensured that an arbitrary state should reach the lower
limit of
 bound (\ref{neqa}) for all $\theta > 0$. But it is sure for an infinitesimal evolution. So bound (\ref{neqa}) is a good estimation
  about an evolution with very short $\tau$ and very small $\theta$.

If a state has the maximum $\Delta E$, the state should have the form $[\vert E_{\rm min}\rangle+\exp(i\phi)\vert E_{\rm max}\rangle]/\sqrt2$,
where $E_{\rm min}$ and $E_{\rm max}$ are the minimal and maximum energies of the system, and $\phi$ is an arbitrary phase. The state
reaches the lower limit of bound (\ref{neqa}) for any $\theta$. But the maximum energy deviation is not necessary for reaching the
lower limit. On the other hand, if a state reaches the lower limit of bound (\ref{neqa}) for any $\theta$, one necessary condition
is that  the state can evolve to an orthogonal state.

%\section{Applications}
{\bf Application.}  We apply the GLM-$\beta$ bound in (\ref{neqa})
to estimate the evolution of quantum gates. The Walsh-Hadamard
transformation is a well-known one-qubit gate, which transform qubit
state $\vert 0 \rangle$ to
  $(\vert 0 \rangle + \vert 1 \rangle)/\sqrt2$ and $\vert 1 \rangle$ to $(\vert 0\rangle -\vert1\rangle)/\sqrt2$.
 We  consider the
   evolution from an initial state $\vert 0\rangle$ to the final state $(\vert 0\rangle+\vert1\rangle)/\sqrt2$ (It is obvious that
    $\theta=\pi/4$ ) in a one-qubit system having a general Hamiltonian
\[
H=-\varepsilon\sigma_3+\delta_1\sigma_1+ \delta_2\sigma_2,
\]
where $\sigma_1,\;\sigma_2,\;\sigma_3$ are the three Pauli matrices.
Let $\delta_1=\delta\cos\phi$, $\delta_2=\delta\sin\phi$ and
$\sin\beta=\delta/\sqrt{\varepsilon^2+\delta^2}$, it is easy to
check that the system has  $\Delta E=\delta$ and the state at an
arbitrary time
\begin{eqnarray}
\vert\psi(t)\rangle &=&(\cos\frac{\sqrt{\varepsilon^2+\delta^2}
t}{\hbar}+ i \cos\beta\sin\frac{\sqrt{\varepsilon^2+\delta^2}
t}{\hbar}) \vert 0\rangle \nonumber\\
&+&  i\mathrm
\exp(i\phi)\sin\beta\sin\frac{\sqrt{\varepsilon^2+\delta^2}
t}{\hbar}\vert1\rangle.
\end{eqnarray}
By selecting a suitable phase $\phi$ (suppose that $\delta_1$ and
$\delta_2$ can be adjusted) and time $\tau$, one can carry out the
 Walsh-Hadamard transformation on the initial state $\vert0\rangle$,
 where the evolution time $\tau$ satisfies
\[
\tan\left(\frac{\sqrt{\varepsilon^2+\delta^2}}{\hbar}\tau\right)=\sqrt{\frac{\delta^2+\varepsilon^2}{\delta^2-\varepsilon^2}}.
\]
It is obvious that the transformation cannot be realized if $\delta
< \varepsilon$. Hence one has the extent of time evolution
\[
\Delta E\cdot\tau=\frac{\hbar\arcsin\sqrt{\frac{1+(\varepsilon/\delta)^2}{2}}}{ \sqrt{1+(\varepsilon/\delta)^2}}\in\left[h/8,\;h/2\sqrt2\right].
\]
If the system has a very large $\delta\gg\varepsilon$, the lower
limit of bound (\ref{neqa}) can be reached asymptotically while the
lower limit of bound (\ref{neqb1}) can not be reached. Larger is
$\delta$, faster is the evolution, irrespective of the detailed
distribution of $\delta_1$ and  $\delta_2$. It is interesting that
bound (\ref{neqb1}) can be tighter when $\delta$ is small, because
\[
\bar E-E_{\rm min}= \sqrt{\varepsilon^2+\delta^2}-\varepsilon \le \Delta E.
\]

Another example  is the  important two-qubit gate---CNOT gate,
$|0\ket\bra 0|\otimes I^2+ |1\ket\bra 1|\otimes \sigma^2_1$, where
the superscripts denote the qubit numbering. The CNOT gate is
important in quantum information.  Assume that the two-qubit system
has a simple intrinsic Hamiltonian
\[
H=-\varepsilon\left(\sigma_3^1+\sigma_3^2-\sigma_3^1\sigma_3^2\right),
\]
where the superscripts denote the qubit numbering. The CNOT gate can
be realized in three steps: (i) apply a Walsh-Hadamard
transformation to the second qubit through using an
 additional Hamiltonian $H_{\rm 2}=\delta\sigma_2^2$ (corresponding a radio frequency pulse along the $x$-axis), where $\delta$
  is much larger so that the intrinsic evolution can be neglected.  (ii) the state evolves with the intrinsic
  Hamiltonian for a period of $\tau_2=h/(8 \varepsilon)$. (iii) apply an inverse Walsh-Hadamard transformation to the second qubit.
  It is obvious that steps (i), (iii) can reach the lower limit of bound (\ref{neqa}) asymptotically, while the evolution of step (ii)
   almost does not. Here one has
\[
\Delta E\cdot\tau=\frac{\sqrt3h}8>\hbar\theta=\frac{h}{6}.
\]
Although the initial state and the final state spans an angle
$\theta=\pi/3$, CNOT gate cannot be realized without additional
operations, because the initial state is an eigenstate of the
intrinsic Hamiltonian. If the system has the
 intrinsic Hamiltonian $H=\varepsilon (I^1-\sigma_3^1)\sigma_2^2$, CNOT gate can be realized through a intrinsic evolution for a period
 of  $\tau=h/(8\varepsilon)$, which does not reach the lower limit of bound (\ref{neqa}), either. One has $\Delta E\cdot\tau=h/4>\hbar\theta=h/6$.

An interesting example is the well-known Grover's quantum searching algorithm. It contains about $\frac\pi4\sqrt{N}$ Grover iterations,
 where $N$ is the dimension of the database and each iteration rotates a state by $2\arcsin \frac1{\sqrt N}$ \cite{Grover, Grover2}.
 If $N$ is large enough, which is common for an actual database, one has $\theta\sim 2/\sqrt N$. Increasing $N\to\infty$, the Grover iteration
 can possibly be replaced by an infinitesimal evolution having $\tau_I=2\hbar/(\Delta E\sqrt N)$ using the bound in Eq. (\ref{neqa}).
 Thus the minimum total time to carry out
 Grover's quantum searching is $\tau \sim h/4\Delta E$. This result is magical: though one needs iterating times in the order of $\sqrt N$,
 the minimum total time has nothing to do with the database size, and  it can even decrease if  $\Delta E$ increases with an increase in $N$.

GLL is supported by the National Natural Science Foundation of China
Grant No. 10775076 and  the SRFPD Program of Education Ministry of
China (20060003048). DL and YSL are supported by the National
Natural Science Foundation of China Grant No. 10874 098, the
National Basic Research Program of China (2006CB921106).


\begin{thebibliography}{99}
\bibitem{Margolus} N. Margolus and L. B. Levitin, Physica (Amsterdam)120D, 188 (1998).
\bibitem{Fleming} G. N. Fleming, Nuovo Cimento A16, 232 (1973).
\bibitem{Anandan} J. Anandan and Y. Aharonov, Phys. Rev. Lett. 65, 1697(1990).
\bibitem{Vaidman} Lev Vaidman, Am. J. Phys. 60, 182 (1992).
\bibitem{bi11} L. B. Levitin, T. Toffoli, and Z. Walton, in Quantum Commmunication, Measurement and Computing, edited by J. Schapiro and
O. Hirota (Rinton, Princeton, 2003), pp. 457¨C459.
\bibitem{bi12} L. Levitin, T. Toffoli, and Z. Walton, Int. J. Theor. Phys. 44, 965 (2005).
\bibitem{bi8} V. Giovannetti, S. Lloyd, and L. Maccone, Phys. Rev. A 67, 052109 (2003).
\bibitem{bi9} V. Giovannetti, S. Lloyd, and L. Maccone, Europhys. Lett. 62, 615 (2003).
\bibitem{bi10} C. Zander, A.R. Plastino, A. Plastino, and M. Casas, J. Phys. A 40, 2861 (2007).
\bibitem{bi13} P. Kosin¡äski and M. Zych, Phys. Rev. A 73, 024303 (2006).
\bibitem{bi14}  M. Andrecut and M. K. Ali, Int. J. Theor. Phys. 43, 969 (2004).
\bibitem{bi15} D. C. Brody, J. Phys. A 36, 5587 (2003).
\bibitem{bi16} B. Zielinski and M. Zych, Phys. Rev. A 74, 034301 (2006).
\bibitem{bi17} D. C. Brody and D.W. Hook, J. Phys. A 39, L167 (2006).
\bibitem{bi18} A. Carlini, A. Hosoya, T. Koike, and Y. Okudaira, Phys. Rev. Lett. 96, 060503 (2006).
\bibitem{bi19} S. Luo, Physica (Amsterdam) 189D, 1 (2004).
\bibitem{bi20} A. K. Pati, Phys. Lett. A 262, 296 (1999).
\bibitem{bi21} P. Pfeifer, Phys. Rev. Lett. 70, 3365 (1993).
\bibitem{bi22} A. Carlini, A. Hosoya, T. Koike, and Y. Okudaira, arXiv: quant-ph/06080039v4.
\bibitem{bi23} A. Borras, C. Zander, A. R. Plastino, M. Casas, and A. Plastino, Europhys. Lett. 81, 30007 (2008).
\bibitem{bi24} J. So¡§derholm, B. Gunnar, T. Tedros, and A. Trifonov, Phys. Rev. A 59, 1788 (1999).
\bibitem{bi25} J. Uffink, Am. J. Phys. 61, 935 (1993).
\bibitem{bi26} S. Lloyd, Nature (London) 406, 1047 (2000).
\bibitem{bi27} S. Lloyd, Phys. Rev. Lett. 88, 237901 (2002).
\bibitem{Levitin} Lev B. Levitin and Tommaso Toffoli, Phys. Rev. Lett.103, 160502(2009).
\bibitem{Grover} L.K. Grover, Phys. Rev. Lett. 79, 325(1997).
\bibitem{Grover2} L.K. Grover, Phys. Rev. Lett. 80, 4329(1998).
\end{thebibliography}
\end{document}